\documentclass[5p,number,times,sort&compress]{elsarticle}
\usepackage{amsmath,graphicx,url,color}

\newcommand*{\half}{\frac{1}{2}}
\newcommand*{\ds}{\displaystyle}
\newcommand*{\D}{\mathrm{d}}
\newcommand*{\I}{\mathrm{i}}
\newcommand*{\Exp}[1]{\mathrm{e}^{\mbox{\footnotesize$#1$}}}
\newcommand*{\ket}[1]{{\left|{#1}\right\rangle}}
\newcommand*{\bra}[1]{{\left\langle{#1}\right|}}  
\newcommand*{\braket}[3][]{{\left\langle{#2}{#1|}{#3}\right\rangle}}
\newcommand*{\expect}[2][]{{\left\langle\vphantom{#1|}{#2}\right\rangle}}
\newcommand*{\prp}{\mbox{\scriptsize$\perp$}}
\newcommand*{\repr}{\mathbin{\hat{=}}}
\newcommand*{\column}[2][c]{{\left(\begin{array}{#1}#2\end{array}\right)}}
\newcommand*{\magn}[1]{%
  \mathopen{\boldsymbol{|}}#1\mathclose{\boldsymbol{|}}}
\newcommand*{\bmagn}[1]{%
  \mathopen{\boldsymbol{\big|}}#1\mathclose{\boldsymbol{\big|}}}
\newcommand*{\Bmagn}[1]{%
  \mathopen{\boldsymbol{\Big|}}#1\mathclose{\boldsymbol{\Big|}}}
\newcommand*{\real}[1]{\Real{\left({#1}\right)}}
\newcommand*{\imag}[1]{\Imag{\left({#1}\right)}}
\newcommand*{\Eq}[1]{Eq.~(\ref{eq:#1})}
\newcommand*{\U}[1]{{\small\bfseries U#1}}
\DeclareMathOperator{\Real}{Re}
\DeclareMathOperator{\Imag}{Im}
\DeclareMathOperator{\var}{var}
\DeclareMathOperator{\ssd}{ssd}
\DeclareMathOperator{\mtc}{mtc}
\DeclareMathOperator{\be}{be}
\DeclareMathOperator{\cor}{cor}
\DeclareMathOperator{\rmI}{I}
\DeclareMathOperator*{\Min}{Min}
\DeclareMathOperator*{\Max}{Max}

\newcommand*{\sfrac}[2]{\mbox{\small$\displaystyle\frac{#1}{#2}$}}
\newcommand*{\pheq}{\mathrel{\phantom{=}}}
\newcommand*{\scalemath}[2]{\scalebox{#1}{$\displaystyle{#2}$}}

\DeclareMathAlphabet{\vecfont}{OT1}{cmr}{bx}{it}
\renewcommand{\vec}[1]{\vecfont{#1}}

\makeatletter%
\newcommand{\ps@arXiv}{%
  \renewcommand{\@evenhead}{}%
  \renewcommand{\@oddhead}{\hfil%
    \raisebox{8ex}{\small(Posted on the arXiv on 8 October 2023,
      v2 on 1 December 2023)}\hfil}%
  \renewcommand{\@oddfoot}{\small{}%
                    \hfil\thepage\hfil}%
  \renewcommand{\@evenfoot}{\@oddfoot}}
\renewcommand{\ps@pprintTitle}{\ps@arXiv}
\makeatother 

\begin{document}

\title{Uncertainty relations revisited$^1$\fnref{dedic}}
\fntext[dedic]{Dedicated to Daniel M. Greenberger on the occasion of his 90th
  birthday.} 

\author{Berthold-Georg~Englert\fnref{email}}
\fntext[email]{\textit{Email address}: \texttt{englert@u.nus.edu}}%
  \affiliation[1]{%
    organization={Key Laboratory of Advanced Optoelectronic Quantum
      Architecture and Measurement of Ministry of Education, \\
      School of Physics, Beijing Institute of Technology}, %
    city={Beijing}, postcode={100081}, country={China}}
  \affiliation[2]{organization={Centre for Quantum Technologies, %
      National University of Singapore}, %
    city={Singapore}, postcode={117543}, country={Singapore}}
  \affiliation[3]{%
    organization={Department of Physics, National University of Singapore}, %
    city={Singapore}, postcode={117542}, country={Singapore}}

\begin{abstract}
Introductory courses on quantum mechanics usually include lectures on
uncertainty relations, typically the inequality derived by Robertson and,
perhaps, other statements.
For the benefit of the lecturers, we present a unified approach ---
well suited for undergraduate teaching --- for deriving all standard
uncertainty relations: those for products of variances by Kennard, Robertson,
and Schr\"odinger, as well as those for sums of variances by Maccone and Pati.  
We also give a brief review of the early history of this topic and try to
answer why the use of variances for quantifying uncertainty is so widespread,
while alternatives are available that can be more natural and more fitting.

It is common to regard the states that saturate the Robertson inequality as
``minimum uncertainty states'' although they do not minimize the variance of
one observable, given the variance of another, incompatible observable.
The states that achieve this objective are different and can be found
systematically.   

\end{abstract}


\maketitle

\section{Introduction}\noindent%
Uncertainty relations in quantum mechanics date back to the late 1920s, when
Heisenberg's 1927 paper \cite{Heisenberg:1927} caught the attention of Kennard
\cite{Kennard:1927}, Robertson \cite{Robertson:1929}, Schr\"odinger
\cite{Schrodinger:1930}, and others.
One could think that this topic was exhausted decades ago but this is not the
case.
For example, there is the 2014 paper by Maccone and Pati
\cite{Maccone+Pati:2014} that supplements the earlier inequalities for
products of variances with inequalities for sums.

In addition, there is the question about the role played by the states for
which the Roberston inequality is saturated, commonly called ``minimum
uncertainty states.''
This is misleading terminology; it should no longer be repeated in lectures
and textbooks because the states that actually minimize the variance of one
observables, given the variance of another, are different.

We address this issue in Sec.~\ref{sec:RS}, after first offering general
remarks about measures of uncertainty, variances and others, in
Sec.~\ref{sec:MoU} and remarks on why the use of variances is so widespread in
Sec.~\ref{sec:WV?}.
Section~\ref{sec:STD} presents a derivation of all standard inequalities for
products and sums of variances from a basic observation; this requires only
mathematical tools that are available to undergraduate students.
Section~\ref{sec:UP} reviews the uncertainty principle, in particular its
historical origin.
Then, in Sec.~\ref{sec:SUS} we consider the actual minimum uncertainty states,
followed by identifying, in Sec.~\ref{sec:CH}, the (lower) convex hull of the
region occupied by the variances of a pair of observables in a two-dimensional
plot. 
The examples in Sec.~\ref{sec:EX} are suitable for undergraduate courses.
Some technical details are reported in the appendixes.

It should be clear that the subject matter of uncertainty relations is more
important for the teaching of quantum mechanics than for research at the
frontier of knowledge.
Accordingly, this paper aims at providing material that, in the first place,
is useful for the learning students and their lecturers.

\section{Measures of uncertainty}\label{sec:MoU}\noindent%
A measure of uncertainty quantifies the fuzziness associated with the
probability distribution of a random variable, such as the values of a
measured property of a quantum system.
While there is no unique way of quantifying uncertainty, all uncertainty
measures must have the following three properties:
\begin{equation}\label{eq:unc1a}
  \begin{tabular}{@{}l@{\quad}p{0.8\columnwidth}@{}}
    \U1 & {\raggedright{}They are well-defined and assign a nonnegative
           number to every distribution of the random variable;}\\ 
    \U2 & {\raggedright{}they acquire the minimal value (usually $=0$) in all
           limits of a sharp distribution (only one value of the random
           variable occurs), and only then;}\\ 
    \U3 & {\raggedright{}they are concave: a convex sum of distributions
           cannot have an uncertainty measure less than the corresponding
           average of the uncertainty measures for the ingredient
           distributions.} 
  \end{tabular}
\end{equation}
There is also a desirable fourth property:
\begin{equation}\label{eq:unc1b}
  \begin{tabular}{@{}l@{\quad}p{0.8\columnwidth}@{}}
    \U4 & {\raggedright{}A measure of uncertainty should
          reach its largest value (finite or infinite)
          in the limit of a uniform distribution.}
  \end{tabular}
\end{equation}
These properties are closely analogous to those required of measures for path
knowledge in interferometers \cite{Durr:2001,Englert+3:2008}.
Note that property \U2 excludes measures that assign minimal uncertainty to
distributions with nonzero probabilities for more than one value; and that the
``should reach'' clause in property \U4 allows measures that are maximal
for nonuniform distributions.  

One important example for which this clause matters is the \emph{variance}
\begin{equation}\label{eq:unc2}
  \var(X)=\Min_{x}\Bigl\{\expect{(X-x)^2}\Bigr\}
  =\expect{X^2}-\expect{X}^2
\end{equation}
for a real-valued random variable $X$; it has properties \U1, \U2, and \U3,
but is usually largest for a nonuniform distribution.
In \Eq{unc2}, all real reference values $x$
participate in the competition.
It is natural to restrict the $x$ values to the proper values of $X$, and then
we obtain another permissible measure, the \emph{smallest squared distance}
\begin{equation}\label{eq:unc3}
  \ssd(X)=\var(X)+\bigl(\expect{X}-\check{x}\bigr)^2\,,
\end{equation}
where $\check{x}$ is the proper value closest to $\expect{X}$.
Many more ways of quantifying uncertainty are available,
and one should choose the uncertainty measure that fits the context.

In the spirit of $\ssd(\ )$, there are, in particular, the measures based on
minimizing a Monge-type  cost, $\mtc(\ )$, for converting the actual
probability distribution into a sharp one,
\begin{equation}\label{eq:unc4}
  \mtc(X)=\Min_{x}\Bigl\{\bigl\langle c(x,X)\bigr\rangle\Bigr\}\,,
\end{equation}
where $c(x,x')$ is the cost associated with allocating the probability of
finding the proper value $x'$ to that of finding $x$; we have
$c(x,x')=(x-x')^2$ in \Eq{unc3}.
The minimization ensures property \U3, and
\begin{equation}\label{eq:ubc5}
  c(x,x')\geq0\quad\mbox{with ``$=$'' if $x=x'$ and only then}
\end{equation}
guarantees property \U2.
This approach was explored by Werner and collaborators, see Refs.\
\cite{Busch+2:2014,Dammeier+2:2015,Werner:2016,Busch+2:2018}
and works cited there.

Yet another procedure favors entropic measures that do not change when the
proper values of $X$ are permuted, such as a
Shannon-type entropy for the probabilities of a discrete observable or the
binned probabilities of a continuous observable
\cite{Bia-Bir+1:1975,Maassen+1:1988,Abdelkhalek+7:2015}.
While entropic measures of this kind possess property \U4 and find applications
in quantum information science (see, e.g., \cite{Coles+3:2017}), they do not
serve the same purpose as (the square root of) the variance or a Monge-type
cost, namely quantifying our uncertainty about, say, the position by a
distance measure.

\section{Why variances?}\label{sec:WV?}\noindent%
It appears that the common practice --- almost without exceptions --- of
measuring an uncertainty in quantum mechanics by a variance originates in a
first-day plausible choice that no one questioned.
In Heisenberg's ground-breaking 1927 paper on how to understand the
then-new quantum mechanics \cite{Heisenberg:1927}, there are two relations
between uncertainty products and Planck's constant $h$,
\begin{equation}\label{eq:var1}
  \text{first:}\ q_1p_1\sim h\,,\quad \text{later:}\ q_1p_1=\frac{h}{2\pi}\,.
\end{equation}
The first is an intuitive, semi-quantitative, order-of-magnitude statement
about ``the precision $q_1$ with which the position is known
and the precision $p_1$ with which the momentum is determinable;''%
\footnote{This is not an exact quote but a paraphrase of Wheeler's and Zurek's
  translation of Heisenberg's words.}
a microscope with its finite resolution and the recoil of the Compton effect
are mentioned in this context.
While this seems to put $q_1$ and $p_1$ on different footing,
the supporting evidence of the later equality has them on equal footing
because Heisenberg employs for this purpose a wave function that gives 
gaussian probability densities ${\propto\!\exp\bigl(-(q-q')^2/q_1^2\bigr)}$ and
${\propto\!\exp\bigl(-(p-p')^2/p_1^2\bigr)}$ for the position $Q$ and the
momentum $P$, respectively.

Heisenberg did not offer a general definition of $q_1$ and $p_1$ for
arbitrary wave functions and seemed to trust that his intuition would tell him
how to quantify uncertainty in any given situation.
What would Heisenberg's 1927 intuition have told him about a position
distribution with two well-separated narrow peaks?
We don't know.
It is, however, not plausible that he would have chosen the variance for
quantifying the position uncertainty in this situation.  

Still in 1927, Kennard worked out the now-common textbook examples for which
Heisenberg's equations of motion are linear and, therefore, easily solved 
\cite{Kennard:1927}.
He mathematized Heisenberg's $q_1p_1$ product by defining both factors as the
square roots of doubled variances and derived the inequality
\begin{equation}\label{eq:var1'}
    q_ip_i\geq \frac{h}{2\pi}\,,
\end{equation}
which states the ``slightly generalized law of indeterminacy by Heisenberg''  
for ``every pair of canonical variables'' 
(analogs of position and momentum).
Except for the doubling, Kennard's definition became the accepted notion of
uncertainty immediately:
Weyl used it in his 1928 book, then Condon, Robertson, and Schr\"odinger in
their papers of 1929 and 1930
\cite{WeylGT+QM:1928,Condon:1929,Robertson:1929,Schrodinger:1930}. 
Since none of these authors justified the use of variances --- Robertson
simply invokes ``statistical usage'' --- this must have
been an utterly natural choice for them and their contemporaries.

While the use of variances for measuring uncertainty can be questioned, there
is a context in which variances appear naturally, namely in the no-change
probability when a small change is induced by a unitary transformation,
\begin{equation}\label{eq:var2}
  \Bmagn{\expect{\Exp{\I\epsilon X}}}^2=1-\epsilon^2\var(X)+\cdots\,,
\end{equation}
where the ellipsis stands for contributions of order $\epsilon^4$ or higher
powers.
None of the classical papers on the subject mentions this, however, and so we
conclude that the use of variances was not motivated by \Eq{var2}.

In the terminology of Ref.~\cite{Dammeier+2:2015}, the inequalities by
Kennard, Robertson, and Schr\"odinger are statements about the
\emph{preparation uncertainty}, not about the
\emph{error-disturbance uncertainty} or the \emph{measurement uncertainty}.
The well-known works by Arthurs and Kelly \cite{Arthurs+1:1965}, and by Ozawa
\cite{Ozawa:1984} concern the latter.
In this paper, we deal only with the preparation uncertainty.

The initial steps taken by Heisenberg and Kennard and followed
by others gave the center stage to inequalities obeyed by the product of two
variances.
Of course, it was clear early on that there can be vanishing variances for
observables with discrete eigenvalues \cite{Condon:1929}, in which case
inequalities for sums of the two variances are more useful.
Judging by the wikipedia article on the topic \cite{wikipedia}, the
inequalities derived by Maccone and Pati in 2014 \cite{Maccone+Pati:2014} are
regarded as standard in this context.

\section{The standard inequalities}\label{sec:STD}\noindent%
The inequalities for products of variances are usually demonstrated as 
consequences of the Cauchy--Bunyakovsky--Schwarz inequality, whereas Maccone and
Pati employed much more advanced tools when establishing their inequalities for
sums.
Contrary to the impression this might create, the two kinds of uncertainty
relations are not profoundly different. 
In fact, all standard inequalities --- for products or sums of
variances --- follow from the elementary fact that an operator maps a ket onto
a linear combination of this ket and a second, perpendicular ket.

We consider a quantum system, whose state is described by the generic
unit-length ket $\ket{\ }$ and its adjoint bra $\bra{\ }$, and two real physical
properties symbolized by the hermitian operators $A$ and $B$.
While the following considerations have easy generalizations for mixed states,
we simplify the presentation by the use of pure states. 
We then have ${\expect{A}=\bra{\ }A\ket{\ }}$, for instance.

For any two complex numbers $a$ and $b$ --- with metrical dimensions such that
$aA+bB$ is meaningful 
--- the identity
\begin{equation}\label{eq:A2}
  (aA+bB)\,\ket{\ }=\ket{\ }\,\bigl(a\expect{A}+b\expect{B}\bigr)
  +\ket{\prp}\,\gamma
\end{equation}
defines the product $\ket{\prp}\,\gamma$ of the unit-length ket $\ket{\prp}$
and the complex coefficient $\gamma$, with $\braket{\ }{\prp}=0$ and
\begin{align}\label{eq:A3}
  \magn{\gamma}^2
  &=\magn{a}^2\var(A)+\magn{b}^2\var(B)\nonumber\\
  &\pheq\mbox{}
    +2\real{a^*b}\cor(A,B)+2\imag{a^*b}\,\expect{\sfrac{\I}{2}[A,B]}\,,
\end{align}
where ${[A,B]=AB-BA}$ is the commutator and
\begin{equation}\label{eq:A4}
  \cor(A,B)=\frac{1}{2}\expect{(AB+BA)}-\expect{A}\expect{B}
\end{equation}
is the correlation of $A$ and $B$.
With judicious choices for $a$ and $b$ and the observation that
\begin{equation}\label{eq:A5}
  \magn{\gamma}^2\geq\bmagn{\bra{\smash{\prp'}}\,(aA+bB)\,\ket{\ }}^2\geq0
\end{equation}
for all unit-length bras $\bra{\prp'}$ that are orthogonal to the state bra
$\bra{\ }$, ${\braket{\prp'}{\ }=0}$,
\Eq{A3} implies the uncertainty relations by Kennard, Robertson,
Schr\"odinger, and Maccone and Pati.
The equal sign holds in the left relation in \Eq{A5} when
${\bra{\prp'}=\bra{\prp}}$. 

The first Maccone--Pati inequality \cite[Eq.~(3)]{Maccone+Pati:2014} is
obtained for ${a^*b=\pm\I\magn{ab}}$ with the sign such that the right-hand
side in 
\begin{align}\label{eq:A6}
  \magn{a}^2\var(A)+\magn{b}^2\var(B)&\geq
  \bmagn{\bra{\smash{\prp'}}\,(aA+bB)\,\ket{\ }}^2\nonumber\\
  &\pheq\mbox{}\mp\magn{ab}\,\expect{\I[A,B]}
\end{align}
is largest.
When ${a^*b=\pm\magn{ab}}$, we first learn that
\begin{equation}\label{eq:A7}
  2\bmagn{ab\cor(A,B)}\leq\magn{a}^2\var(A)+\magn{b}^2\var(B)
\end{equation}
and then infer the second  Maccone--Pati inequality
\cite[Eq.~(4)]{Maccone+Pati:2014},
\begin{equation}\label{eq:A8}
  \magn{a}^2\var(A)+\magn{b}^2\var(B)
  \geq\frac{1}{2}\bmagn{\bra{\prp}\,(aA+bB)\,\ket{\ }}^2\,.
\end{equation}
Since \Eq{A7} holds for all $\magn{a}$, $\magn{b}$, the more stringent
inequality
\begin{equation}\label{eq:A9}
  \cor(A,B)^2\leq\var(A)\var(B)
\end{equation}
is true and establishes a third inequality of the Maccone--Pati kind,
\begin{equation}\label{eq:A10}
  \magn{a}\var(A)^{1/2}+\magn{b}\var(B)^{1/2}
  \geq\bmagn{\bra{\prp}\,(aA+bB)\,\ket{\ }}\,.
\end{equation}

Next, we write ${a^*b=\magn{ab}\,\Exp{\I\phi}}$, bound the $\phi$-dependent
$\magn{\gamma}^2$ by ${\magn{\gamma}^2\geq0}$, then 
minimize the right-hand side in \Eq{A3} over the phase $\phi$,
\begin{align}\label{eq:A11}
  0&\leq\magn{a}^2\var(A)+\magn{b}^2\var(B)\nonumber\\
  &\pheq\mbox{}
    -2\magn{ab}\,\Biggl(\cor(A,B)^2
                      +\expect{\sfrac{\I}{2}[A,B]}^{\!2}\Biggr)^{1/2}\,,
\end{align}
and conclude that the Schr\"odinger inequality holds
\cite[Eq.~(9)]{Schrodinger:1930}, 
\begin{equation}\label{eq:A12}
 \var(A)\var(B)\geq\cor(A,B)^2+\expect{\sfrac{\I}{2}[A,B]}^2\,.
\end{equation}
When we discard the correlation term, or choose ${a^*b=\pm\I\magn{ab}}$
to begin with, we have the Robertson inequality
\cite[4th eq.]{Robertson:1929},
\begin{equation}\label{eq:A13}
  \var(A)\var(B)\geq\expect{\sfrac{\I}{2}[A,B]}^2\,.
\end{equation}
We apply this to a Heisenberg pair of
observables, such as position ${A=Q}$ and momentum ${B=P}$ with
${\I[Q,P]=-\hbar}$, and arrive at the Kennard inequality
\cite[Eq.~(27)]{Kennard:1927}, 
\begin{equation}\label{eq:A14}
  \var(Q)\var(P)\geq\frac{1}{4}\hbar^2\,.
\end{equation}
Further, we get \Eq{A9} by discarding the commutator term in \Eq{A12},
or by choosing ${a^*b=\pm\magn{ab}}$ to begin with.

One can also establish the Robertson and Schr\"odinger inequalities by
combining the ${b=0}$ and ${a=0}$ versions of Eqs.~(\ref{eq:A2}) and
(\ref{eq:A3}) \cite{Goldenberg+1:1996}.
More inequalities result from yet other choices for $a$ and $b$; we
leave it at that.

We note in passing that the Schr\"odinger inequality is not stronger
than the Robertson inequality
--- although wikipedia, e.g., says so \cite{wikipedia} ---
because they imply each other.
In particular, we obtain \Eq{A12} by the replacements
\begin{equation}\label{eq:A15}
  A\to\lambda^{-1}A+\lambda B\,,\quad B\to\lambda^{-1}A-\lambda B\,,
\end{equation}
in \Eq{A13} with ${\lambda^2=\sqrt{\var(A)/\var(B)}}\,$.
We shall not explore the stronger-weaker angle any further.

\section{Uncertainty principle}\label{sec:UP}\noindent%
Heisenberg never spoke of the \emph{uncertainty principle},%
\footnote{See, e.g.,
  Note~1 in \cite{Sen:2014} or Sec.~2.4 in \cite{Hilgevoord+Uffink:2016}.}
Kennard didn't use the term either but referred to \Eq{A14} as
the ``law of indeterminacy'' (\textsl{Unbestimmtheitsgesetz}
\cite[p.~339]{Kennard:1927})  
while others used ``uncertainty principle'' quite early --- and the term stuck.
In his 1929 paper \cite{Condon:1929}, Condon remarked on ``uncertainty
principles'' (plural) and, without referring to Kennard's work, he stated that
\Eq{A14} ``has come to be known quite generally as Heisenberg's
uncertainty relation''  --- the terms  ``uncertainty
principles'' and ``uncertainty relations'' were synonyms for Condon. 
Robertson's way of using ``uncertainty principle'' later in 1929
\cite{Robertson:1929} suggests that this was established terminology, and he
sneaked it into the index of the English translation of Weyl's book
\cite{WeylGT+QMeng:1931}.  
This index entry points to the page where the Kennard inequality
is stated and to the appendix where it is demonstrated  but ``uncertainty
principle'' is nowhere in Weyl's text;
the German editions of 1928 and 1931 \cite{WeylGT+QM:1928,WeylGT+QMzw:1931} do
not have such an index entry.
Incidentally, just like Condon and Robertson, Weyl did not mention Kennard's
contribution, he acknowledged Pauli.  

\begin{figure}[t]
  \centering
  \includegraphics[viewport=90 570 262 742,clip=]{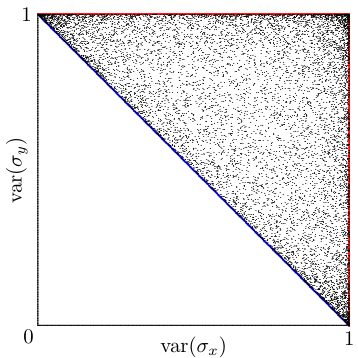}
  \caption{\label{fig:1a}%
    Variances of two Pauli operators.
    The scattered dots mark pairs of variances for a Haar-uniform sample of
    pure qubit states.
    The blue line is the border identified by the uncertainty relation in
    \Eq{B2a}. 
    The red lines show the variances of the Robertson states.}
\end{figure}

For us, the inequalities obeyed by uncertainty measures are
\emph{uncertainty relations}, and we use the term ``uncertainty principle''
with the limited, and perhaps rudimentary, meaning that there are no physical
pairs of uncertainty values in the vicinity of the bottom-left corner
in plots such as the ones in Figs.~\ref{fig:1a}--\ref{fig:5} in the following
sections if the uncertainties refer to incompatible observables ---
observables that do not have one or more common eigenstates.
This is in line with Heisenberg's profound insight that, as a matter of
principle, one cannot have precise knowledge of two such observables:
We can know either one of the properties with great precision ---
with a small uncertainty or no uncertainty at all, that is ---
but there is no situation in which we know both properties accurately on the
quantum scale.

Having thus clarified what the ``uncertainty principle'' means for us, we
submit that this is poorly chosen terminology as it is not a
physical principle in the usual sense.
While the principle of relativity, the entropy principle in statistical physics,
the principle of complementarity in quantum kinematics,
the equivalence principle in general relativity, the gauge principle in
quantum field theory, and the various action principles are the starting
points of whole branches of physics or impose profound constraints, the
uncertainty principle plays no such role.  

The standard inequalities in \mbox{Sec.~\ref{sec:STD}}, taken together, serve
the purpose of confirming that the uncertainty principle is respected by each
pair of incompatible observables if we measure uncertainties by variances.
When other measures are used, corresponding inequalities apply.

\section{Robertson states}\label{sec:RS}\noindent%
The Robertson inequality in \Eq{A13} is textbook fare, often called ``the
general uncertainty principle'' or similar; see, e.g., Sec.~24 in
\cite{GottfriedQM:1966}. 
It is saturated if 
\begin{equation}\label{eq:B1}
  (aA+bB)\,\ket{\ }=\ket{\ }\,\bigl(a\expect{A}+b\expect{B}\bigr)
  \ \text{with}\ \real{a^*b}=0\,. 
\end{equation}
States described by such kets are commonly termed ``minimum uncertainty
states'' (see, e.g., Ch.~8, Sec.~6 in \cite{MerzbacherQM:1961}),
which alleges that they solve a minimization problem, namely that of finding
the smallest value of $\var(A)$ for a given value of $\var(B)$ or
\textsl{vice versa}.
This is not actually the case and, therefore, we use the neutral term
``Robertson states'' for the states defined by \Eq{B1}.

Figure \ref{fig:1a} illustrates this point for two Pauli operators of a qubit,
such as ${A=\sigma_x}$ and ${B=\sigma_y}$ with
${\var(\sigma_x)=1-\expect{\sigma_x}^2}$, for instance; see \ref{app:qubit}.
We have
\begin{equation}\label{eq:B2a}
  \var(\sigma_x)+\var(\sigma_y)\geq1\,,
\end{equation}
for all pairs of variances.
The equal sign holds for all pure qubit states with
${\expect{\sigma_z}=0}$; these are the true minimum-uncertainty states here.
The Robertson states, however, are pure states with either
${\var(\sigma_x)=1}$ or ${\var(\sigma_y)=1}$.

\begin{figure}[t]
  \centering
  \includegraphics[viewport=87 561 258 732,clip=]{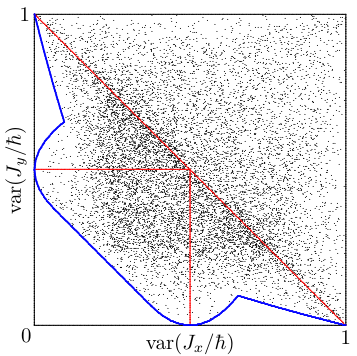}
  \caption{\label{fig:2}%
    Variances of two angular-momentum component for spin one.
    The scattered dots mark pairs of variances for a Haar-uniform sample of
    pure states.
    The blue line is the border identified in \Eq{E4}. 
    The red lines show the variances of the Robertson states.
    See also Fig.~4 in \cite{Dammeier+2:2015} and Fig.~3 in
    \cite{Szymanski+1:2020}.} 
\end{figure}

A second example are the variances of two components of the angular momentum
vector $\vec{J}$ for spin one, shown in Fig.~\ref{fig:2}.
Clearly, the variances of the Robertson states (red lines) do not
inform us about the border (blue); see \ref{app:spin1}.

\section{Stationary uncertainty states}\label{sec:SUS}\noindent%
Figures \ref{fig:1a} and \ref{fig:2} prompt us to ask
\begin{equation}\label{eq:C1}
    \parbox[b]{0.81\columnwidth}{%
        \centerline{\underline{The central question}}\rule{0pt}{2ex}%
    For two hermitian observables, $A$ and $B$, what are the values
    available for the pairs of their variances?}
\end{equation}
A partial answer is given by finding the smallest value of $\var(A)$ for a
given value of $\var(B)$ and \textsl{vice versa}. 
It is obvious that the Robertson states are useless here;
in fact, it is unclear which purpose they serve.%
\footnote{El Kinani and Daoud call the Robertson states ``intelligent states''
  and use them for a construction of overcomplete sets of kets and bras
  \cite{Kinani+1:2001}; I owe this remark to Magdalena Zych.
  See \cite{Mista+3:2022} for a recent variation on this theme.} 

The full answer to the central question requires knowledge of the
\emph{stationary uncertainty states} which we find in reply to 
\begin{equation}\label{eq:C2}
  \parbox[b]{0.81\columnwidth}{%
 \centerline{\underline{The auxiliary questions}}\rule{0pt}{2ex}%
 (i)\;\ For the variances of two hermitian observables,
 what are the stationary pairs
 of their values?\newline\rule{0pt}{2ex}
  (ii)\;\ What are the states for which one variance is stationary, given the
  value of the other?}
\end{equation}
The familiar tools of variational calculus establish that the auxiliary
question (ii) is answered by the states that solve the nonlinear eigenvalue
equation 
\begin{equation}\label{eq:C3}
   \Bigl(\alpha\bigl(A-\expect{A}\bigr)^2
  +\beta\bigl(B-\expect{B}\bigr)^2\Bigr)\,\ket{\ }
  =\ket{\ }\,\bigl(\alpha\var(A)+\beta\var(B)\bigr) 
\end{equation}
where $\alpha$ and $\beta$ are real Lagrange parameters.
The variances associated with these states provide the answer to the auxiliary
question (i); among them are the pairs on the border of the accessible region,
the answer to the central question.

\section{Convex hull}\label{sec:CH}\noindent%
The smallest eigenvalues for $\alpha,\beta>0$ in \Eq{C3} provide
\begin{equation}\label{eq:D1}
  u(\alpha,\beta)=\Min_{\rho}\bigl\{\alpha\var(A)+\beta\var(B)\bigr\}\,,
\end{equation}
where the minimization is over all statistical operators $\rho$.
We note that, owing to property \U3, it is enough to consider pure states
${\rho=\ket{\ }\bra{\ }}$ for the minimization in \Eq{D1};
and that all pairs $(x,y)$ with ${\alpha x+\beta y\ge u(\alpha,\beta)}$ are in
the (lower) convex hull of the region occupied by the physical
${(x,y)=\bigl(\var(A),\var(B)\bigr)}$ pairs.

More specifically, a pair $(x,y)$ is on the border of the convex hull if
\begin{equation}\label{eq:D2}
  x=\Max_{\alpha,\beta>0}
  \biggl\{\frac{u(\alpha,\beta)-\beta y}{\alpha}\biggr\}
\quad\mbox{or}\quad
  y=\Max_{\alpha,\beta>0}
    \biggl\{\frac{u(\alpha,\beta)-\alpha x}{\beta}\biggr\}\,.
\end{equation}
Since $u(\alpha,\beta)$ is homogeneous,
\begin{equation}\label{eq:D3}
  u(\alpha,\beta)=\alpha u(1,\beta/\alpha)=\beta u(\alpha/\beta,1)\,,
\end{equation}
\Eq{D2} are essentially statements of a Legendre transformation if
the partial derivatives are defined.
If so, then
\begin{equation}\label{eq:D4}
  x=\frac{\partial u(\alpha,\beta)}{\partial\alpha}\,,\quad
  y=\frac{\partial u(\alpha,\beta)}{\partial\beta}
\end{equation}
give a pair on the border of the convex hull.
Alternatively, one recognizes an application of the Hellmann--Feynman theorem
in \Eq{D4}.

We further note that establishing ${u(\alpha,\beta)>0}$ for just one
$\alpha,\beta$ pair implies that there is a finite-size rectangular triangle
with the right angle at ${(x,y)=(0,0)}$ that contains no
$\bigl(\var(A),\var(B)\bigr)$ pairs.
This demonstrates the uncertainty principle (in the sense of Sec.~\ref{sec:UP})
for the $A,B$ pair under consideration.

\section{Examples}\label{sec:EX}\noindent\vspace*{-2.0\baselineskip}%
\subsection{Position and momentum --- a singularity}\noindent%
For the Heisenberg pair of ${(A,B)=(Q,P)}$, there is no difference between
$\bigl(\var(Q),\var(P)\bigr)$ and $\bigl(\ssd(Q),\ssd(P)\bigr)$.
Equation (\ref{eq:C3}) is the eigenvalue equation of a harmonic oscillator,
and we find 
\begin{equation}\label{eq:E1}
  \var(Q)\var(P)=\frac{1}{4}\hbar^2(2n+1)^2\quad\mbox{with $n=0,1,2,\dots$}
\end{equation}
for the stationary values.
The minimization in \Eq{D1} yields
\begin{equation}\label{eq:E2}
  u(\alpha,\beta)=\hbar\sqrt{\alpha\beta}\,.
\end{equation}
The physical region is convex here; its border is also the border of the convex
hull. 
On the border, we have ${n=0}$ in \Eq{E1} and saturate the Kennard inequality
(\ref{eq:A14}).
This is the singular situation in which the Robertson inequality is the
Kennard inequality, and the Robertson states identify the border.

To appreciate this singularity, we note that the ket of a Robertson state, a
solution of \Eq{B1}, obeys 
\begin{equation}\label{eq:E3}
  \Bigl(\magn{a}^2\bigl(A-\expect{A}\bigr)^2
  +\magn{b}^2\bigl(B-\expect{B}\bigr)^2\Bigr)\,\ket{\ }
  =-\I[A,B]\,\ket{\ }\imag{a^*b}\,,
\end{equation}
which differs in structure from \Eq{C3} unless the commutator $\I[A,B]$ is a
number (multiple of the identity).
This is the case only for Heisenberg pairs of the position-momentum kind.

\begin{figure}[t]
  \centering
  \includegraphics[viewport=90 570 262 742,clip=]{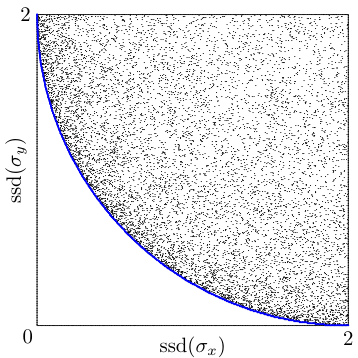}
  \caption{\label{fig:1b}%
    Smallest squared distances of two Pauli operators.
    The scattered dots mark pairs of smallest squared distances
    for a Haar-uniform sample of pure qubit states.
    The blue quarter circle is the border identified by the uncertainty
    relation in \Eq{B2b}.}
\end{figure}

\subsection{Qubit}\noindent%
We supplement the uncertainty relation in \Eq{B2a} for the variances of
$\sigma_x$ and $\sigma_y$ with that for their smallest squared distances,
\begin{equation}\label{eq:B2b}
  \bigl(2-\ssd(\sigma_x)\bigr)^2+\bigl(2-\ssd(\sigma_y)\bigr)^2\leq4\,,
\end{equation}
which is an immediate consequence of
\begin{equation}\label{eq:B3}
  \ssd(\sigma_x)=2-2\,\magn{\expect{\sigma_x}}\,,\quad
  \ssd(\sigma_y)=2-2\,\magn{\expect{\smash{\sigma_y}}}\,.
 \end{equation}
 The matters are illustrated in Fig.~\ref{fig:1b}.

 The permissible pairs of ${(x,y)=\bigl(\var(\sigma_x),\var(\sigma_y)\bigr)}$ in
Fig.~\ref{fig:1a} and of  ${(x,y)=\bigl(\ssd(\sigma_x),\ssd(\sigma_y)\bigr)}$
in Fig.~\ref{fig:1b} cover convex sets.
They are identified by
\begin{equation}\label{eq:B4a}
  u(\alpha,\beta)=\Min\{\alpha,\beta\}
\end{equation}
and
\begin{equation}\label{eq:B4b}
  u(\alpha,\beta)=2(\alpha+\beta)-2\sqrt{\alpha^2+\beta^2}\,,
\end{equation}
respectively.

\subsection{Spin one (variances)}\noindent%
For the spin-one example in Fig.~\ref{fig:2}, stationary
uncertainty states of three kinds make up pieces of the blue border.
The corresponding pairs of variances
${(x,y)=\bigl(\var(J_x/\hbar),\var(J_y/\hbar)\bigr)}$ trace out
\begin{align}\label{eq:E4}
  &\mbox{the straight line $\ds x+y=\frac{7}{16}$ with
    $\ds\magn{x-y}\leq\frac{5}{16}\,;$} \nonumber\\
  &\mbox{the parabola $\ds y=(2x-1)^2$ with $y\leq1\,;$}\nonumber\\
  &\mbox{the parabola $\ds x=(2y-1)^2$ with $x\leq1\,.$}  
\end{align}
All of the straight line is one segment of the border, and each parabola
contributes two segments; see \ref{app:spin1}.
We have
\begin{equation}\label{eq:E5}
  u(\alpha,\beta)=\frac{8\alpha\beta-\Min\{\alpha,\beta\}^2}
                       {16\Max\{\alpha,\beta\}}
\end{equation}
here and obtain the border between ${(x,y)=(0,\half)}$ and $(\half,0)$ by
means of Eqs.~(\ref{eq:D2}).

There are also stationary uncertainty states whose pairs of variances reside
inside the accessible region; the points on the diagonal ${x+y=1}$ are of this
sort.
By an accidental coincidence, the pairs of variances on this diagonal are also
obtained for certain Robertson states, which are not stationary uncertainty
states.%
\footnote{Many other interesting insights about expectation values of powers
  of angular momentum operators, not limited to ${j=\half}$ or ${j=1}$, are
  reported by Sehrawat \cite{Arun:2020}.}

\begin{figure}[t]
  \centering
  \includegraphics[viewport=87 561 258 732,clip=]{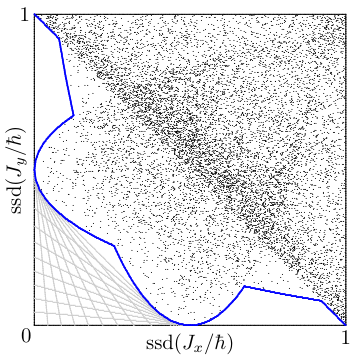}
  \caption{\label{fig:3}%
    Smallest squared distances of two angular-momentum components for spin one.
    The scattered dots mark pairs of smallest squared distances for a
    Haar-uniform sample of pure states. 
    The blue line is the border specified in \Eq{E6};
    the straight gray lines are those of \Eq{E8}.}
\end{figure}

\subsection{Spin one (smallest squared distances)}\label{sec:qubit-ssd}\noindent%
Figure~\ref{fig:3} shows the situation for the smallest squared distances for
the spin-1 pair of ${A=J_x/\hbar}$ and ${B=J_y/\hbar}$.
On the left-hand side in \Eq{C3}, the expectation values of $A$ and $B$ are
replaced by the nearest eigenvalues (here: $0,\pm1$), and we have $\ssd(A)$
and $\ssd(B)$ in the place of $\var(A)$ and $\var(B)$, respectively, in
Eqs.~(\ref{eq:C3}) and (\ref{eq:D1}).

Here, too, we have stationary uncertainty states of three kinds that
make up pieces of the blue border.
For $(x,y)=\bigl(\ssd(J_x/\hbar),\ssd(J_y/\hbar)\bigr)$ there are 
\begin{align}\label{eq:E6}
  &\mbox{two straight-line segments $\ds x+y=1\,;$}\nonumber\\
  &\mbox{two elliptical arcs $\ds(2x-1)^2+(1-y/2)^2=1\,;$}\nonumber\\
  &\mbox{two elliptical arcs $\ds(2y-1)^2+(1-x/2)^2=1\,.$}  
\end{align}
The border of the convex hull between ${(x,y)=(0,\half)}$ and $(\half,0)$
results from 
\begin{align}\label{eq:E7}
  u(\alpha,\beta)&=\half\Min\{\alpha,\beta\}+2\Max\{\alpha,\beta\}
                   \nonumber\\&\pheq
  -\half\sqrt{\Min\{\alpha,\beta\}^2+16\Max\{\alpha,\beta\}^2}\,.
\end{align}
In accordance with Sec.~\ref{sec:CH}, the straight lines
\begin{equation}\label{eq:E8}
  \alpha x+\beta y=u(\alpha,\beta)
\end{equation}
are tangential to the border and identify the convex hull;
see the gray lines in Fig.~\ref{fig:3}.

\subsection{Qutrit --- complementary pair of unitary observables}%
\label{sec:qutrit-ssd}\noindent%
Two components of the angular momentum vector operator are not complementary
for spin~1 (or any higher value).
Instead, a natural pair of complementary observables for the qutrit are Weyl's
period-$3$ unitary operators $U$ and $V$ \cite{WeylGT+QMeng:1931,Schwinger:1960}
with the matrix representations
\begin{equation}\label{eq:E9}
  U\repr\column[ccc]{1&0&0\\0&\ \Exp{\I2\pi/3}&0\\0&0&\Exp{-\I2\pi/3}}\,,\quad
  V\repr\column[ccc]{0&1&0\\0&0&1\\1&0&0}\,.
\end{equation}
Their eigenvalues $1$, $\Exp{\I2\pi/3}$, $\Exp{-\I2\pi/3}$ are the corners of
an equilateral triangle in the complex plane.
We use the sides of the triangle to quantify the distance between the corners.
The smallest squared distance is then proportional to the sum of the two
smallest probabilities in the trio $p_1,p_2,p_3$ with
\begin{equation}\label{eq:E10}
  p_k=\expect{\delta\Bigl(U,\Exp{\I(2\pi/3)k}\Bigr)}
\end{equation}
for $U$, that is
\begin{equation}\label{eq:E11}
  \ssd(U)=\frac{3}{2}-\frac{3}{2}
  \Max_k\biggl\{\expect{\delta\Bigl(U,\Exp{\I(2\pi/3)k}\Bigr)}\biggr\}\,,
\end{equation}
and analogously for $V$; the factor of $\frac{3}{2}$ is for normalization:
${0\leq\ssd(U)\leq1}$.

\begin{figure}[t]
  \centering
  \includegraphics[viewport=87 561 258 732,clip=]{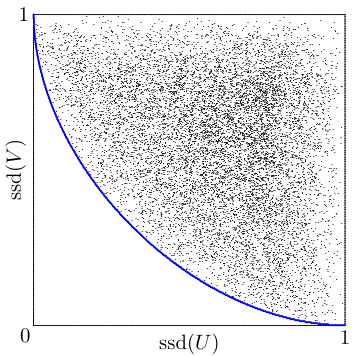}
  \caption{\label{fig:4}%
    Smallest squared distances of the Weyl pair of unitary period-$3$ operators
    for a qutrit.
    The scattered dots mark pairs of smallest squared distances
    for a Haar-uniform sample of pure states.
    The blue line is the border in \Eq{E13}.}
\end{figure}

The physical set of ${(x,y)=\bigl(\ssd(U),\ssd(V)\bigr)}$ pairs is convex, and
we find the border from
\begin{align}\label{eq:E12}
  u(\alpha,\beta)
  &=\frac{3}{4}(\alpha+\beta)
    -\frac{3}{4}\sqrt{\alpha^2+\beta^2-\frac{2}{3}\alpha\beta}\,.
\end{align}
The blue border line in Fig.~\ref{fig:4} is an arc of the ellipse
\begin{align}\label{eq:E13}
  \frac{1}{3}(2x+2y-3)^2+\frac{2}{3}(x-y)^2=1\,;
\end{align}
see Sec.~4.1.1 in \cite{Englert+3:2008}.

\subsection{Period-$N$ Weyl pairs}\noindent%
More generally than the period-2 pair $\sigma_x,\sigma_y$ for the qubit and
the period-3 pair $U,V$ for the qutrit, we have the Weyl pair of complementary
period-$N$ unitary operators $U,V$ for a $N$-dimensional degree of freedom
\cite{WeylGT+QMeng:1931,Schwinger:1960}.
The two orthonormal eigenstate bases are unbiased,
\begin{equation}\label{eq:W1}
  \braket{u_k}{v_l}=\frac{1}{\sqrt{N}}\Exp{\I2\pi kl/N}\,,
\end{equation}
where ${u_k=\Exp{\I2\pi k/N}}$ and ${v_l=\Exp{\I2\pi l/N}}$ are the eigenvalues
with ${k,l=0,1,2\dots,N-1}$.
Then
\begin{equation}\label{eq:W2}
  c\bigl(u_k,u_{k'}\bigr)=\frac{N}{N-1}\bigl(1-\delta_{k,k'}\bigr)
  \ \ \mbox{and}\ \ 
  c\bigl(v_l,v_{l'}\bigr)=\frac{N}{N-1}\bigl(1-\delta_{l,l'}\bigr)
\end{equation}
specify particularly simple Monge-type costs,
\begin{equation}\label{eq:W3}
  \mtc(U)=\frac{N}{N-1}\Bigl(1-\Max_k\Bigl\{\expect{\delta(U,u_k)}\Bigr\}\Bigr)
\end{equation}
and analogously for $\mtc(V)$.
This is an example of an uncertainty measure that reaches its largest value
for the uniform distribution only. 
For ${N=2}$, this gives $\half\ssd(\sigma_x)$ and $\half\ssd(\sigma_y)$ of
Sec.~\ref{sec:qubit-ssd};
for ${N=3}$, it gives $\ssd(U)$ and $\ssd(V)$ of Sec.~\ref{sec:qutrit-ssd}.
For any $N$ value, we have a convex set of physical
$(x,y)=\bigl(\mtc(U),\mtc(V)\bigr)$ pairs, whose border is identified by
\begin{equation}\label{eq:W4}
  u(\alpha,\beta)=\half\frac{N}{N-1}
  {\left(\alpha+\beta-\sqrt{\alpha^2+\beta^2-\frac{2N-4}{N}\alpha\beta}
      \,\right)}\,.
\end{equation}
This border is the arc of the ellipse
\begin{equation}\label{eq:W5}
  \frac{(N-1)^2}{N}\biggl(\frac{N}{N-1}-x-y\biggr)^2+\frac{N-1}{N}(x-y)^2=1
\end{equation}
that connects $(x,y)=(0,1)$ and $(x,y)=(1,0)$; see Sec.~6.2 in
\cite{Englert+3:2008}. 
For $N=2$, it is the quarter circle ${(1-x)^2+(1-y)^2=1}$ (cf.\
Fig.~\ref{fig:1b}); in the limit ${N\to\infty}$, it is the straight line
${x+y=1}$.

\subsection{Rotor variables}\noindent%
The quantum rotor has a periodic position (azimuth) with bras
${\bra{\varphi}=\bra{\varphi+2\pi}}$
that we picture as points on a circle, and a discrete momentum (orbital
angular momentum for the symmetry axis of the circle) with kets $\ket{\ell}$
for ${\ell=0,\pm1,\pm2,\dots}$;
the continuous $\varphi$ basis and the discrete $\ell$ basis are unbiased
\cite[Sec.~1.1.9]{Durt+3:2010}, 
\begin{equation}\label{eq:E14}
  \braket{\varphi}{\ell}=\Exp{\I\ell\varphi}\,.
\end{equation}
We associate a unitary observable $E$ with the quantum number $\varphi$ and a
hermitian observable $L$ with the quantum number $\ell$,
\begin{equation}\label{eq:E15}
  \bra{\varphi}E=\Exp{\I\varphi}\bra{\varphi}\,,\quad
  L\ket{\ell}=\ket{\ell}\ell\,.
\end{equation}
We use $\ssd(L)$ of \Eq{unc3} for $L$ and, in analogy with
Sec.~\ref{sec:qutrit-ssd}, we measure the uncertainty of $E$ by the smallest
squared distance that goes  with the chordal distance between points on the
circle, 
\begin{equation}\label{eq:E16}
  \ssd(E)=\Min_{\varphi}\Bigl\{1-\real{\expect{E}\Exp{-\I\varphi}}\Bigr\}
  =1-\magn{\expect{E}}\,,
\end{equation}
scaled such that ${0\leq\ssd(E)\leq1}$.
This is a Monge-type cost with
${c\Bigl(\Exp{\I\varphi},\Exp{\I\varphi'}\Bigr)=1-\cos(\varphi-\varphi')}$ in
\Eq{unc4};
it is another example of an uncertainty measure that is largest for
the uniform distribution, but there are other distributions as well for which
${\expect{E}=0}$. 

\begin{figure}[t]
  \centering
   \includegraphics[viewport=85 560 260 735,clip=]{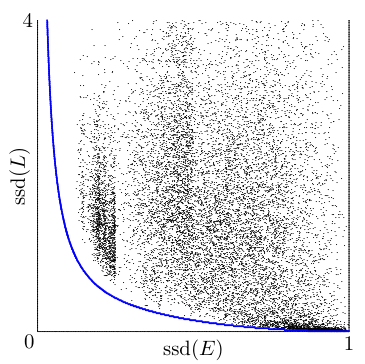}
  \caption{\label{fig:5}%
    Smallest squared distances of the rotor variables.
    The scattered dots that mark pairs of smallest squared distances
    are for several samples of pure states. 
    The blue line is the border resulting from \Eq{E18}.}
\end{figure}

The stationary uncertainty states are solutions of
\begin{align}\label{eq:E17}
  &\pheq\Biggl(\alpha\biggl(1-\scalemath{0.8}{\half}
    \bigl(E^{\dagger}\Exp{\I\phi}
  +E\Exp{-\I\phi}\bigr)\biggr)
  +\beta(L-l)^2\Biggr)\,\ket{\ }\nonumber\\
  &=\ket{\ }\,\bigl(\alpha\ssd(E)+\beta\ssd(L)\bigr)\,,
\end{align}
where ${\expect{E}=\magn{\expect{E}}\,\Exp{\I\phi}}$ and $l$ is the integer
nearest to $\expect{L}$.
The periodic wave function
${\psi(\varphi)=\braket{\varphi}{\ }=\psi(\varphi+2\pi)}$ obeys a
Mathieu-type equation (see \ref{app:rotor}), from which we obtain
\begin{equation}\label{eq:E18}
  u(\alpha,\beta)=\frac{1}{4}\beta\be_0(8\alpha/\beta)\,,
\end{equation}
where $\be_0(\ )$ is the zeroth characteristic value of the Mathieu equation
with the notational conventions of \cite{NBS-Mathieu:1951}.
This gives the blue border of the convex set of the physical
${\bigl(\ssd(E),\ssd(L)\bigr)}$ pairs in Fig.~\ref{fig:5}.
The other stationary values are located on the lines associated with
\begin{equation}\label{eq:E19}
  u_n(\alpha,\beta)=\frac{1}{4}\beta\be_n(8\alpha/\beta)
  \quad\mbox{with $n=1,2,3,\dots$}
\end{equation}
in analogy with Eqs.~(\ref{eq:D2}) and (\ref{eq:D4}).
Similar to \Eq{E1}, these lines do not intersect.

\section{Summary}\noindent%
We recalled the early history of uncertainty relations, and traced the
widespread use of variances as measures of uncertainty to Kennard's
mathematization of Heisenberg's observation.
The unified approach by which we derived all standard uncertainty relations
can be taught in undergraduate courses.

Further, we noted that calling the states that saturate Robertson's inequality
``minimum uncertainty states'' is misleading as the challenge of minimizing
the variance of one observable, given the variance of another, is met by other
states.
We also discussed other measures of uncertainty, in particular the smallest
squared distance and the Monge-type cost.
The position-momentum pair of linear motion, the qubit,
the spin-1 angular momentum, the qutrit, the Weyl pairs
of period-$N$ observables, and the rotor provided illustrating examples,
which --- so we hope --- will be useful when teaching quantum mechanics.

\section{Acknowledgments}\noindent%
I wish to thank Reinhard Werner and Zden\v{e}k Hradil for wakening my dormant 
interest in this topic, and Jos\'e Ignacio Latorre for helpful discussions.
This work was supported by the National Research Foundation, Singapore and
A*STAR under its CQT Bridging Grant.

\appendix\renewcommand{\thefigure}{\arabic{figure}}
\section{Qubit}\label{app:qubit}\noindent%
The uncertainty relations in Eqs.~(\ref{eq:B2a}) and (\ref{eq:B2b}) follow
from
\begin{equation}\label{eq:app-1}
  \expect{\sigma_x}^2+\expect{\smash{\sigma_y}}^2\leq
  \expect{\sigma_x}^2+\expect{\smash{\sigma_y}}^2+\expect{\sigma_z}^2\leq1\,.  
\end{equation}
Equations (\ref{eq:B4a}) and (\ref{eq:B4b}) are other immediate consequences
of \Eq{app-1}.

The Robertson states solve
\begin{equation}\label{eq:app-2}
  \bigl(a\sigma_x+\I b\sigma_y\bigr)\,\ket{\ }
  =\ket{\ }\,\Bigl(a\expect{\sigma_x}+\I b\expect{\smash{\sigma_y}}\Bigr)
\end{equation}
with real $a$ and $b$. Since the two eigenvalues of $a\sigma_x+\I b\sigma_y$
are either both real (when $a^2>b^2$) or both imaginary (when $a^2<b^2$), we
have either $\expect{\smash{\sigma_y}}=0$ or $\expect{\sigma_x}=0$ in a
Robertson state and, therefore, either $\var(\sigma_y)=1$ or
$\var(\sigma_x)=1$.  

\section{Spin one}\label{app:spin1}\noindent%
We use the eigenkets of $J_x$, $J_y$, and $J_z$ with eigenvalue $0$ for the
orthonormal basis in the ket space, and represent $J_x$ and $J_y$ with real
symmetric matrices, 
\begin{equation}\label{eq:app-3}
  \frac{1}{\hbar}J_x\repr\column[crr]{0&0&0\\ 0&0&1\\ 0&1&0}\,,\quad
  \frac{1}{\hbar}J_y\repr\column[rcr]{0&0&1\\ 0&0&0\\ 1&0&0}
\end{equation}
There are three families of Robertson states with the kets represented by the
columns 
\begin{equation}\label{eq:app-4}
  \column[@{}c@{}]{\cos(\phi)\\ \I\sin(\phi)\\ 0}\,,\ 
  \frac{1/\sqrt{2}}{\cosh(\theta)}
    \column[@{}c@{}]{\I\sinh(\theta)\\ \cosh(\theta)\\ \pm1}\,,\ 
  \frac{1/\sqrt{2}}{\cosh(\theta)}
    \column[@{}c@{}]{\cosh(\theta)\\\I\sinh(\theta)\\ \pm1}\,,
\end{equation}
where $\phi$ and $\theta$ are real parameters.
The respective $(x,y)=\bigl(\var(J_x/\hbar),\var(J_y/\hbar)\bigr)$ pairs are
\begin{equation}\label{eq:app-5}
  (x,y)=\left\{\begin{array}{l}
       \Bigl(\sin(\phi)^2,\cos(\phi)^2\Bigr)\,,\\[1.5ex]
 \ds\half\Bigl(\tanh(\theta)^2,1\Bigr)\,,\\[2ex]
  \ds\half\Bigl(1,\tanh(\theta)^2\Bigr)\,.\end{array}\right.  
\end{equation}
They trace out the red lines in Fig.~\ref{fig:2}. \clearpage

Owing to the choice of matrices in \Eq{app-3}, the solutions of \Eq{C3} for
${A=J_x/\hbar}$ and ${B=J_y/\hbar}$ have kets represented by three-component
columns with real probability amplitudes,
\begin{equation}\label{eq:app-6}
  \ket{\ }\repr\column[@{}c@{}]{\psi_x \\ \psi_y \\ \psi_z}\quad
  \mbox{with}\quad \psi_x^2+\psi_y^2+\psi_z^2=1\,,
\end{equation}
for which
\begin{align}\label{eq:app-7}
  \expect{J_x/\hbar}&=2\psi_y\psi_z\,,\quad
           \var(J_x/\hbar)=\psi_y^2+\psi_z^2-4\psi_y^2\psi_z^2\,,
           \nonumber\\
  \expect{\smash{J_y}/\hbar}&=2\psi_z\psi_x\,,\quad
            \var(J_y/\hbar)=\psi_z^2+\psi_x^2-4\psi_z^2\psi_x^2\,,
\end{align}
and
\begin{align}\label{eq:app-8}
  \ssd(J_x/\hbar)&=2-\psi_x^2-\Max\bigl\{1,\magn{4\psi_y\psi_z}\bigr\}\,,
                   \nonumber\\
  \ssd(J_y/\hbar)&=2-\psi_y^2-\Max\bigl\{1,\magn{4\psi_z\psi_x}\bigr\}\,.
\end{align}
By coincidence, the variances on the line $x+y=1$ are stationary, namely for
\begin{equation}\label{eq:app-9}
  \ket{\ }\repr\column[@{}c@{}]{\cos(\phi) \\ \sin(\phi) \\ 0}\,,
\end{equation}
which are \emph{not} the kets of the first family of Robertson states in
\Eq{app-4}.  
The stationary uncertainty states on the straight-line segment in \Eq{E4} are
\begin{equation}
  \label{eq:app-10}
  \ket{\ }\repr\column[@{}c@{}]{\sqrt{5/8}\,\cos(\phi) \\
    \sqrt{5/8}\,\sin(\phi) \\ \sqrt{3/8}}   
\end{equation}
and those for the parabolic arcs are
\begin{equation}\label{eq:app-11}
  \ket{\ }\repr\column[@{}c@{}]{\cos(\phi) \\ 0 \\ \sin(\phi)}
  \quad\mbox{and}\quad
  \ket{\ }\repr\column[@{}c@{}]{0 \\ \cos(\phi) \\ \sin(\phi)}  \,.
\end{equation}
Now turning to the smallest squared distances in Fig.~\ref{fig:3} and
\Eq{E6}, the straight-line segments are obtained for the kets in \Eq{app-9}
and the elliptical arcs for the kets in \Eq{app-11} with
${\magn{\sin(2\phi)}\geq\half}$.

\section{Rotor}\label{app:rotor}\noindent%
Since
\begin{equation}\label{eq:app-12}
  E\Exp{-\I\phi}=\Exp{-\I\phi L} E \Exp{\I\phi L}\,,\quad
  L-l=E^l L E^{-l}
\end{equation}
it is enough to consider \Eq{E17} for ${\phi=0}$ and ${l=0}$, where it is
essential that $l$ is an integer.
Then
\begin{equation}\label{eq:app-13}
  \pheq\Biggl(\alpha\,\bigl(1-\cos(\varphi)\bigr)
  -\beta\frac{\partial^2}{\partial\varphi^2}\Biggr)\,\psi(\varphi)
  =\psi(\varphi)\,\bigl(\alpha\ssd(E)+\beta\ssd(L)\bigr)  
\end{equation}
is obeyed by $\psi(\varphi)=\braket{\varphi}{\ }=\psi(\varphi+2\pi)$, with the
constraints
\begin{equation}\label{eq:app-14}
  \expect{E}>0\quad\mbox{and}\quad{}-\half<\expect{L}<\half\,,
\end{equation}
that is,\par
\begin{align}\label{eq:app-15}
  &\int\limits_{(2\pi)}\frac{\D\varphi}{2\pi}\,\magn{\psi(\varphi)}^2
  \Exp{\I\varphi}>0\,,\nonumber\\\mbox{and}\quad
  -\half<&\int\limits_{(2\pi)}\frac{\D\varphi}{2\pi}\,
  \psi(\varphi)^*\frac{1}{\I}\frac{\partial}{\partial\varphi}\psi(\varphi)
  <\half\,,
\end{align}
where the integration covers any $2\pi$-interval.

Equation (\ref{eq:app-13}) is a standard Mathieu equation
\cite{NBS-Mathieu:1951},
\begin{equation}\label{eq:app-16}
  y''(x)+\Bigl(b-s\cos(x)^2\Bigr)\,y(x)=0\,,
\end{equation}
with $x=\half\varphi$, $y(x)=\psi(\varphi)$, and
\begin{equation}\label{eq:app-17}
  a=-\frac{8\alpha}{\beta}\,,\quad
  b=-\frac{8\alpha}{\beta}+\frac{4\alpha}{\beta}\ssd(E)+4\ssd(L)\,.
\end{equation}
The smallest characteristic value ${\be_0(s)=s+\be_0(-s)}$ for the even
periodic solutions, ${y(x+\pi)=y(x)=y(-x)}$, tells us about
${u(\alpha,\beta)=\alpha\ssd(E)+\beta\ssd(L)}$ for ${b=\be_0(s)}$,
and so we arrive at $u(\alpha,\beta)$ in \Eq{E18};
see also Sec.~VI\,A\,2 in \cite{Busch+2:2018}.

The points ${(x,y)=\bigl(\ssd(E),\ssd(L)\bigr)}$ with
\begin{align}\label{eq:app-18}
  x&=\frac{\partial u(\alpha,\beta)}{\partial\alpha}=2\be_0'(8\alpha/\beta)
  \,,\nonumber\\
  y&=\frac{\partial u(\alpha,\beta)}{\partial\beta}=
  \frac{1}{4}\be_0(8\alpha/\beta)-2\frac{\alpha}{\beta}\be_0'(8\alpha/\beta)
\end{align}
are on the blue border in Fig.~\ref{fig:5}.
We have
\begin{equation}
\label{eq:app-19}
\begin{array}[b]{l@{\ \mbox{when}\ }l}
  \ds y\simeq\frac{1}{8x}-\frac{1+x}{16} & 0<x\ll1\,,\\[2ex]
  \ds x\simeq1-\sqrt{2y}                 & 0<y\ll1\,,
\end{array}  
\end{equation}
when $x=\ssd(E)$ or $y=\ssd(L)$ is very small, respectively.

We note that \Eq{E17} with $l$ replaced by the noninteger $\expect{L}$ was
also considered by Hradil \textsl{et al.} \cite{Hradil+4:2006} who used
$\var(L)$ to quantify the uncertainty of the angular momentum and
${1-\magn{\expect{E}}^2}$ for that of the azimuth, which is the sum
$\var(C)+\var(S)$ for the hermitian $C$ and $S$ in ${E=C+\I S}$.
Their focus was on a Robertson-type inequality for the product of these
measures of uncertainty.

\begin{figure}[t]
  \centering
  \includegraphics[viewport=75 568 270 740,clip=]{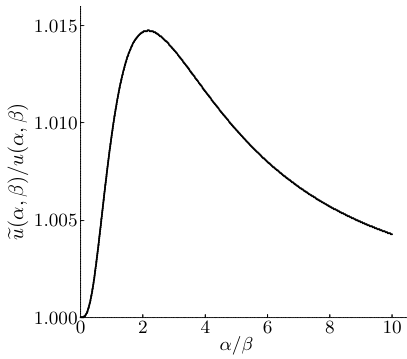}
  \caption{\label{fig:7}%
    The ratio of $\,\widetilde{u}(\alpha,\beta)$ in \Eq{app-22} and
    $u(\alpha,\beta)$ in \Eq{E18} offers a comparison of the uncertainties
    for the von Mises states with those of the
    minimum-uncertainty states.}
\end{figure}

It is noteworthy that the pairs of uncertainties
\begin{equation}\label{eq:app-20}
  \ssd(E)=1-\frac{\rmI_1(t)}{\rmI_0(t)}=X(t)\,,\quad
  \ssd(L)=\frac{t\rmI_1(t)}{4\rmI_0(t)}=Y(t)\,, 
\end{equation}
are very close to the blue border in Fig.~\ref{fig:5};
here, $\rmI_0(\ )$ and $\rmI_1(\ )$ are standard modified Bessel functions and
$t\geq0$. 
These uncertainties are the smallest squared distances for the states with the
kets 
\begin{equation}\label{eq:app-21}
  \ket{\phi,\ell,t}
  =\Exp{-\I(L-\ell)\phi}\Exp{Ct}\ket{\ell}\Big/\!\sqrt{\rmI_0(t)}\,,
\end{equation}
which are called von Mises states because they yield a von Mises distribution
\cite{Mises:1918} for the azimuth; they are Robertson states of some sort
\cite{Mista+3:2022}. 
To demonstrate the case, we first note that the asymptotic forms in
\Eq{app-19} are also correct for the pair in \Eq{app-20}, and then
show the graph of the ratio of
\begin{align}\label{eq:app-22}
  &\widetilde{u}(\alpha,\beta)=\alpha X(t)+\beta Y(t)\nonumber\\
  \mbox{with}\quad
  &\alpha \frac{\D X(t)}{\D t}+\beta\frac{\D Y(t)}{\D t}=0\,,
\end{align}
which gives \Eq{app-20} in accordance with \Eq{D4},
and $u(\alpha,\beta)$ in \Eq{E18} in Fig.~\ref{fig:7}.
The ratio exceeds unity, of course, but not by much; it is less than $1.0148$
for all $\alpha/\beta$ and the largest near $\alpha/\beta=2.164$.


\end{document}